\def\editmode{0}

\def\bibfilenames{bibman_refs,refs}

\def\spsformat{0}

\if\spsformat1

\documentclass{article}
\usepackage{spconf}

\else
\documentclass[10pt,conference]{IEEEtran}

\IEEEoverridecommandlockouts

\fi

\usepackage{include_v7}

\usepackage{notation}

\newcommand{\cbr}[1]{} 

\usepackage{hyperref}
\usepackage{balance}

\begin{document}

\title{Spoofing Attack Detection in the Physical Layer\\
    with Robustness to User Movement\thanks{
        This research has been funded in part by the Research Council of Norway under IKTPLUSS grant 311994.}
    \thanks{The datasets and code are available at \url{https://github.com/fachu000/spoofing}.}
}

\if\spsformat1
    \name{Daniel Romero$^1$, Tien Ngoc Ha $^1$, and Peter Gerstoft$^2$\thanks{
            This research has been funded in part by the Research Council of Norway under IKTPLUSS grant 311994 and Intelligence Advanced Research Projects Activity (IARPA), via 2021-2106240007.}
        \thanks{Data and code to reproduce
            the experiments are available at \url{https://github.com/fachu000/spoofing}}}
    \address{$^1$Dept. of ICT, University of Agder, Norway.  \{daniel.romero,tien.n.ha\}@uia.no \\
        $^2$Electrical and Computer Engineering, University of California, San
        Diego, USA. pgerstoft@ucsd.edu  }
\else
    \author{
        \IEEEauthorblockN{Daniel Romero}
        \IEEEauthorblockA{\textit{
                Department of ICT},
            \textit{University of Agder}\\
            Grimstad, Norway \\
            daniel.romero@uia.no}
        \and
        \IEEEauthorblockN{Tien Ngoc Ha}
        \IEEEauthorblockA{\textit{
                Department of ICT},
            \textit{University of Agder}\\
            Grimstad, Norway \\
            tien.n.ha@uia.no}
        \and
        \IEEEauthorblockN{Peter Gerstoft}
        \IEEEauthorblockA{
            \textit{UC San Diego}\\
            San Diego, USA \\
            pgerstoft@ucsd.edu}

    }

\fi

\maketitle

\begin{abstract}
    In a spoofing attack,  an attacker impersonates a legitimate user to access
    or modify  data belonging to the latter. Typical approaches for spoofing
    detection in the physical layer declare an attack when a change is observed
    in  certain channel features, such as the received signal
    strength (RSS) measured by spatially distributed receivers.
    However, since channels  change over time, for
    example due to user movement,  such  approaches are impractical. To
    sidestep this limitation, this paper proposes a scheme that combines the decisions
    of a position-change detector based on a deep neural network to
    distinguish spoofing from movement. Building upon  community
    detection on graphs, the sequence of received frames is partitioned  into
    subsequences to detect concurrent transmissions from distinct locations. The
    scheme can be easily deployed in practice since it just involves
    collecting a small dataset of  measurements at a few tens of locations that need not even be computed or recorded. The scheme is evaluated on real data collected for this purpose.



\end{abstract}


\section{Introduction}
\label{sec:intro}

\begin{bullets}
    \blt[Motivation]
    \begin{bullets}
        \blt[General]The shared nature of the wireless communication channel
        opens the door for a large number of attacks that seek to illicitly
        obtain private data, compromise the uptime of remote services, or
        impersonate other
        users~\cite{kolias2015intrusion,vanhoef2016mac,martin2019handoff,tippenhauer2011requirements}.
        \blt[spoofing]Spoofing attacks, which pursue the latter goal, are
        especially problematic since they endow the attacker with the capacity
        to access and modify data intended for or produced by a legitimate
        user.
        \blt[detecting spoofing] Detecting these attacks is therefore vital to guarantee data security.

        \blt[PHY]Cryptographic techniques can be
        employed at one or multiple communication layers to preempt spoofing attacks. However, these
        security barriers are readily bypassed by an attacker with the user
        credentials. For this reason, techniques to detect spoofing attacks in
        the physical layer have been developed, typically by exploiting physical
        layer characteristics that are specific to the transmitter, such as the
        carrier frequency offset (CFO), or channel features, such as the
        received signal strength (RSS) or Angle of Arrival (AoA).\cbr{ These
            spoofing detection techniques generally match observed features to
            previously collected ones for a given legitimate transmitter.}
    \end{bullets}%

    \blt[Literature]%
    \begin{bullets}%
        \blt[imperfections]More specifically,\cbr{imperfections such as} CFO, I/Q
        offset, and I/Q imbalance\cbr{, which uniquely characterize the analog
            hardware of transmitters,} were used to verify user identity in
        \cite{brik2008wireless,givehchian2022evaluating,liu2019real,vo2016fingerprinting}. Unfortunately,
        these techniques necessitate knowledge of the communication protocol and may fail under changes in the environment; e.g. in
        temperature~\cite{givehchian2022evaluating}.
        \blt[AoA, TDoA, SNR]These limitations have been partially alleviated in
        \cite{xiong2010secureangle,xiong2013securearray,shi2016robust}, which
        employ AoA and TDoA features, and in~\cite{wang2020machine}, where a
        neural network is trained on signal-to-noise ratio (SNR) traces\cbr{
            obtained in the sector-level sweep process in mm-Wave 60 GHz IEEE
            802.11ad Networks}. However, they still require synchronization and/or
        knowledge of the communication protocol.

        \blt[RSS]In comparison, RSS-based techniques  need not know the
        protocols or decode signals, which greatly enhances their generality and
        applicability for spoofing
        detection~\cite{chen2007detecting,yang2012detection,xiao2016phy,zeng2011identity,alotaibi2016new}.
        The most common technique in this context  applies K-means to RSS
        measurements collected by multiple
        receivers~\cite{chen2007detecting,hoang2018soft,sobehy2020csi}. Leveraging the dependence of RSS signatures on
        the location of the transmitter,  an attack is declared iff
        transmissions with the same user ID\cbr{ (as provided by higher communication
            layers)} originate at different locations.
        This means that attacks are declared \emph{even when the
            channel changes}, which occurs e.g. when a user moves.

        Fortunately, it is possible to  tell spoofing from motion and other
        changes by looking at the nature of the changes. To see this, suppose
        that a network sequentially receives frames from points A, B, C, and D.
        If all these points are different, it is natural to ascribe these changes to user
        movement. On the contrary, if the transmissions are received from points
        A, B, A, B, A, B, etc., it is likely that  one user is transmitting from  A and another
        from B, thereby suggesting the presence of an attack.
        \blt[limitations]However, to our knowledge, no
        existing algorithm leverages this kind of information.

    \end{bullets}

    \blt[contribution] This paper fills this gap by developing
    an RSS-based spoofing attack detector that builds upon the aforementioned
    observation. The main contribution is a two-tier scheme where the decisions
    of a low-level \emph{position-change detector} (PCD), such as the detectors
    alluded to earlier, are combined by a higher-level algorithm to detect the
    presence of concurrent transmissions from different locations. To simplify
    the terminology, the exposition assumes that all changes are due to user
    movement, but the algorithm carries over without modification to
    changes  caused by other factors such as  the movement of
    scatterers or obstructions. The high-level algorithm invokes a graph community detection primitive to
    decompose the received frame sequence into subsets corresponding to the same
    location. Since existing PCDs necessitate accurate RSS estimates, which  requires long averaging time, a secondary contribution is a
    \cbr{data-driven}PCD based on a deep neural network (DNN) that provides a higher spatiotemporal
    resolution by learning the distribution of coarse RSS estimates.

    Collecting data for deploying the proposed algorithm
    is remarkably simple: it just involves recording samples for a few tens of
    transmitter locations. As opposed to related algorithms, such as those for
    fingerprinting localization,  locations need not be computed or
    recorded, which greatly simplifies the process.

\end{bullets}


\section{Model and Problem Formulation}
\label{sec:mpf}

\cmt{model}
\begin{bullets}
    \blt[Space]Let $\Area\subset \rfield^3$ comprise the  coordinates of
    all points in the  spatial region of interest, where  both
    legitimate users and attackers are located.
    \blt[Feature vectors]%
    \begin{bullets}%
        \blt[RSS as a function of position]A transmitter at
        $\Loc\in\Area$
        \begin{bullets}
            \blt[tx. signal] transmits a signal with complex baseband representation  $\TxSig(t)$, where $t$ denotes time.  This signal,
            modeled as an unknown wide-sense stationary stochastic process,
            \blt[rx. signal]is received by $\NumFea$ static receivers, which may
            be e.g. access points or base stations. After downconversion,
            the signal at the
            $\IndFea$th receiver is
            \begin{align}
                \RxSig\NotFea{\IndFea}(\Loc,t) = \Cha\NotFea{\IndFea}(\Loc,t)\ast \TxSig(t) + \Noi\NotFea{\IndFea}(t),
            \end{align}
            where $\ast$ denotes convolution, $\Cha\NotFea{\IndFea}(\Loc,t)$ is
            the unknown deterministic impulse response of the
            bandpass equivalent channel  between  $\Loc$ and the
            $\IndFea$th receiver, and $\Noi\NotFea{\IndFea}(t)$ is noise.
            \blt[RSS]As usual, $\Noi\NotFea{\IndFea}(t)$ is assumed wide-sense stationary and
            uncorrelated with $\TxSig(t)$.  Thus, one can define
            the \emph{received signal strength} (RSS)  $
                \Rss\NotFea{\IndFea}(\Loc) \define 
                \expected{|
                    \RxSig\NotFea{\IndFea}(\Loc,t)
                    |^2},$
            where $\expected$ denotes expectation.
        \end{bullets}%
        \blt[RSS estimates]To estimate $\Rss\NotFea{\IndFea}(\Loc)$, the
        $\IndFea$th receiver averages the square magnitude of $\NumSam$ samples $\{|\RxSig\NotFea{\IndFea}(\Loc,\IndSam \SamPer)|^2\}_{\IndSam}$,
        where $\SamPer$ is the sampling interval. If
        $\RxSig\NotFea{\IndFea}(\Loc,\IndSam \SamPer)$ is ergodic for each $\Loc$,  it follows that $\EstRss\NotFea{\IndFea}(\Loc)$
        converges to $\Rss\NotFea{\IndFea}(\Loc)$ as $\NumSam\rightarrow
            \infty$. This number of averaged samples $\NumSam$ will play an important role in Sec.~\ref{sec:pcd}.

        \blt[RSS vector]A fusion center  forms the RSS
        vector estimate $\EstVecRss(\Loc)\define
            [\EstRss\NotFea{0}(\Loc),\ldots,\EstRss\NotFea{\NumFea-1}(\Loc)]\transpose$
        by gathering these $\NumFea$ RSS estimates. It is assumed that the
        position $\Loc$ of the transmitter does not change significantly
        during the process of acquiring  these estimates, which
        is a mild assumption if $\NumSam$ is small, as considered here.
    \end{bullets}%
    \blt[Frames]%
    \begin{bullets}%
        \blt[frame matrix]Repeating this operation for each of a collection of
        $\NumFra$ consecutive received frames
        results in the frame sequence
        $\EstMatRss\define[\EstVecRss\NotFra{0},\ldots,\EstVecRss\NotFra{\NumFra-1}]$,
        where $\EstVecRss\NotFra{\IndFra}\define \EstVecRss(\Loc\NotFra{\IndFra})$ and
        $\Loc\NotFra{\IndFra}\in \Area$ is the location of the user
        that transmitted the $\IndFra$th frame at the  transmission time.

        \blt[user indices]Out of these $\NumFra$ frames, some belong to the
        legitimate user whereas others may belong to the attacker.
        \begin{bullets}
            \blt[legitimate] Let $\SetIndLeg\subset \{0,\ldots,\NumFra-1\}$
            contain the indices of the  frames transmitted by  the legitimate
            user and
            \blt[attacker]let $\SetIndAtt\define\{0,\ldots,\NumFra-1\}-\SetIndLeg$
            contain the indices of the  frames transmitted by the attacker.
        \end{bullets}%
    \end{bullets}%
\end{bullets}%
\cmt{problem formulation}
\begin{bullets}%
    \blt[Given] Given $\EstMatRss$, the problem is to
    \blt[Decide]decide between the following hypotheses:
    \begin{align}
        \label{eq:maintest}
        \begin{cases}
            \HzSD:~\SetIndAtt=\emptyset
            \\
            \HoSD:~\SetIndAtt\neq\emptyset.
        \end{cases}
    \end{align}
    \blt[Dataset]To assist in this task, a data set of feature vectors
    $\{\EstVecRss\NotEst{\IndEst}(\Loc_\IndDat),$ ${\IndDat=0\ldots
                \NumDat-1},~\IndEst=0,\ldots,\NumEst-1\}$ is given, where
    $\Loc_\IndDat\neq \Loc_{\IndDat'}$ for all $\IndDat \neq \IndDat'$ and
    $\{\EstVecRss\NotEst{\IndEst}(\Loc_\IndDat)\}_{\IndEst=0}^{\NumEst-1}$ denote $\NumEst$ estimates of
    $\VecRss(\Loc_\IndDat)\define [
            \Rss\NotFea{0}(\Loc_\IndDat),\ldots,\Rss\NotFea{\NumFea-1}(\Loc_\IndDat)]\transpose$.
\end{bullets}

\section{DNN-based Spoofing Detector}
\label{sec:proposed}

\cmt{Solution overview}
\begin{bullets}
    \blt[assumption]Clearly, if each $\EstVecRss\NotFra{\IndFra}$ is
    transmitted from a totally different location, possibly far from
    $\EstVecRss\NotFra{\IndFra-1}$ and $\EstVecRss\NotFra{\IndFra+1}$, then
    there is no hope of reasonably solving \eqref{eq:maintest} since the possible
    distributions of $\EstMatRss$ are the same under $\HzSD$ and $\HoSD$.
    Thus, the following
    assumption is required:
    \begin{align}
        \nonumber
        \text{(As1) The frame rate is high relative to the speed of the users.}
    \end{align}
    In other words, the movement of the users between two consecutive frames is
    not too large, or, more precisely,  $\Loc\NotFra{\IndFra}$ is
    close to $\Loc\NotFra{\IndFra-1}$ if both $\IndFra-1$ and $\IndFra$ are in
    either $\SetIndLeg$ or $\SetIndAtt$.

    \begin{figure}[t]
        \centering
        \includegraphics[width=0.45\textwidth]{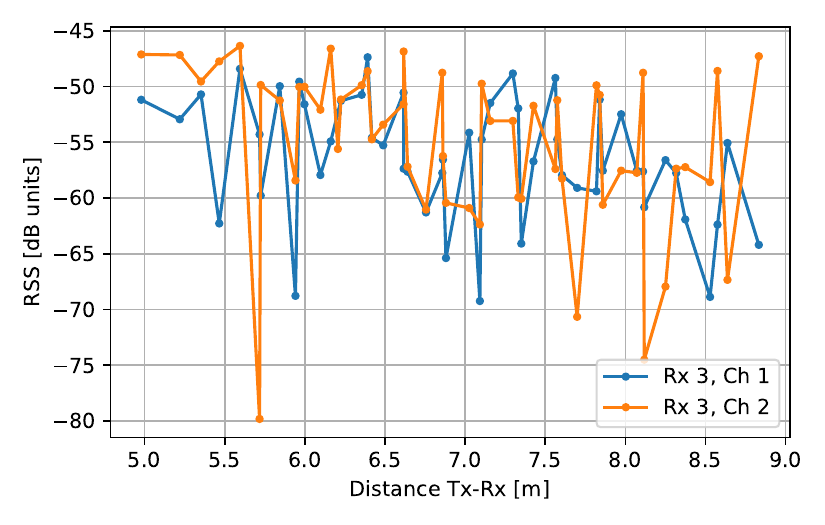}
        \caption{RSS vs. distance for two of the 16 channels in the dataset. It is observed that small distance changes may result in larger RSS changes than large distance changes. }
        \label{fig:fading}
    \end{figure}

    \blt[initial approach]Based on this assumption, one could attempt to solve
    \eqref{eq:maintest} by  first estimating $\Loc\NotFra{\IndFra}$ from
    $\EstVecRss\NotFra{\IndFra}$ for each $\IndFra$ and then deciding that two
    consecutive frames belong to the same user if the corresponding location estimates  are nearby. However, this approach is not viable because (i), as
    per the problem formulation in Sec.~\ref{sec:mpf}, the locations are not
    part of the data,  and (ii), even if the locations were given, estimating
    $\Loc\NotFra{\IndFra}$ entails a large error. This is seen from
    Fig.~\ref{fig:fading}, which shows  $\EstRss\NotFea{\IndFea}(\Loc)$ obtained
    with a large $\NumSam$ for two values of $\IndFea$  vs. the distance from $\Loc$ to the $\IndFea$th receiver in the setup of Sec.~\ref{sec:experiments}. It is seen that small distance changes often  result in larger RSS changes than
    large distance changes. In other words, for this room, the  multipath variability
    in the channel is larger than the path-loss variability. Consequently, the error in estimating $\Loc\NotFra{\IndFra}$ from
    $\EstVecRss\NotFra{\IndFra}$ can be seen to be in the order of the room size, which renders
    such estimates useless. Estimating $\Loc\NotFra{\IndFra}$ with a sufficient
    accuracy would require
    more complex features.

    \blt[proposed approach]Instead, this paper relies on a PCD which decides whether each pair of   frames were
    transmitted from the same location or not,  a much coarser form of
    information than $\Loc\NotFra{\IndFra}$. Based on the PCD decisions,
    a \emph{spoofing detector} (SD) decides $\HzSD$ or $\HoSD$.
    \blt[overview]Although existing algorithms can be used as PCDs,
    Sec.~\ref{sec:pcd}  proposes one especially attuned to low values of
    $\NumSam$. Afterwards, Sec.~\ref{sec:pd} presents the proposed SD.

\end{bullets}

\subsection{Position-Change Detector}
\label{sec:pcd}
\cmt{Overview} This section considers the  subproblem of determining
whether two frames were transmitted from the same location.
\cmt{subproblem}%
\begin{bullets}%
    \blt[Given]Specifically, given
    the estimates $\EstVecRss(\Loc_1)$ and $\EstVecRss(\Loc_2)$    corresponding
    to the (potentially equal) locations $\Loc_1,\Loc_2\in \Area$, the
    problem is to
    \blt[Decide]decide between the following hypotheses:
    \begin{align}
        \label{eq:test}
        \begin{cases}
            \HzPCD:~\Loc_1=\Loc_2
            \\
            \HoPCD:~\Loc_1\neq \Loc_2.
        \end{cases}
    \end{align}
    \blt[Test] A  PCD is a function $\DecPCD:\rfield^{\NumFea} \times
        \rfield^{\NumFea} \rightarrow \{\HzPCD,\HoPCD\}$ that returns  one of these
    hypotheses for each pair $(\EstVecRss,\EstVecRss')$.

    Existing algorithms (cf. Sec.~\ref{sec:intro}) assume a sufficiently large~$\NumSam$, which limits temporal resolution. In contrast, this section proposes a PCD that affords a low
    $\NumSam$ by using   a DNN that  learns the distribution of the feature vector \emph{estimates}.

\end{bullets}

\subsubsection{Data Set}
\label{sec:data}

\begin{bullets}

    \blt[Form pairs]The training dataset is constructed to comprise $2
        \NumPai$ pairs  $ \{
        (\EstVecRss\NotEst{\IndEst_\IndPai}(\Loc_{\IndDat_\IndPai}),
        \EstVecRss\NotEst{\IndEst'_\IndPai}(\Loc_{\IndDat'_\IndPai}))\}_{\IndPai=0}
        ^{2\NumPai-1}\subset\rfield^{\NumFea} \times \rfield^{\NumFea}$. For the first $\NumPai$ pairs, ${\IndDat_\IndPai}={\IndDat'_\IndPai}$ (so $\HzPCD$ holds), whereas for the last $\NumPai$ pairs, ${\IndDat_\IndPai}\neq{\IndDat'_\IndPai}$ (so $\HoPCD$ holds).
    For the $p$th pair, $\IndPai=0,\ldots,2\NumPai-1$, the   indices $\IndEst_\IndPai$ and
    $\IndEst'_\IndPai$ are drawn uniformly at random without replacement from $\{0,\ldots,\NumEst-1\}$.
    For $\IndPai=0,\ldots,\NumPai-1$, index  $ \IndDat_\IndPai=\IndDat'_\IndPai$ is drawn uniformly at random from $\{0,\ldots,\NumDat-1\}$.
    For $\IndPai=\NumPai,\ldots,2\NumPai-1$, indices  $ \IndDat_\IndPai$ and $\IndDat'_\IndPai$ are drawn uniformly at random without replacement from $\{0,\ldots,\NumDat-1\}$.

\end{bullets}

\subsubsection{Architecture}
\begin{bullets}
    \blt[Overview]

    \blt[Test statistic]Detectors are typically designed to decide  $\HoPCD$
    if $\Net(\EstVecRss,\EstVecRss')$ exceeds a  threshold and $\HzPCD$
    otherwise, where function
    $\Net:\rfield^{\NumFea} \times \rfield^{\NumFea} \rightarrow \rfield$ is a
    \emph{detection statistic}. This can be directly implemented as a DNN.

    \blt[DNN]
    \begin{bullets}
        \blt[Commutativization]However, to boost detection performance, it is worth
        applying a technique to make the detector commutative. Specifically, if $\Net$ is just a generic DNN,
        it will generally occur that
        $\NetWei(\EstVecRss,\EstVecRss')\neq\NetWei(\EstVecRss',\EstVecRss)$,
        which is inconsistent with the symmetry of \eqref{eq:test}.
        To remedy this, the proposed PCD uses a DNN to implement an auxiliary function $\AuxNetWei$ and
        then sets
        $
            \NetWei(\EstVecRss,\EstVecRss') = ({
                    \AuxNetWei(\EstVecRss,\EstVecRss') + \AuxNetWei(\EstVecRss',\EstVecRss)
                })/{2}$.
        It can be  seen that
        $\NetWei(\EstVecRss,\EstVecRss')=\NetWei(\EstVecRss',\EstVecRss)$ and
        that the 2 in the denominator can be absorbed into
        $\AuxNetWei(\EstVecRss,\EstVecRss')$.

        \blt[Each component]Function $\AuxNetWei$ is  implemented as a
        feed-forward network whose first layer is  non-trainable and produces
        $10 \log_{10}[\EstVecRss,\EstVecRss',\EstVecRss-\EstVecRss']$ when the network input
        is $[\EstVecRss,\EstVecRss']$. The subsequent 3 hidden layers are fully
        connected with leaky ReLU activations and 512
        neurons~\cite{goodfellow2016}. The output layer is fully connected
        with a single neuron with  linear activation.



    \end{bullets}

\end{bullets}

\subsubsection{Training}
\label{sec:training}

The DNN is trained using a binary cross-entropy loss with $\ell_1$-regularization. A subset of $\NumDat\Val$ points are set apart for validation
and the remaining $\NumDat\Tra\define \NumDat-\NumDat\Val$  are used for
training. The procedure in Sec.~\ref{sec:data} is applied separately to each  subsets by setting $\NumDat$ and $\NumPai$ respectively to $\NumDat\Tra$ and $\NumPai\Tra$ for training and to $\NumDat\Val$ and $\NumPai\Val$ for validation.

\subsection{Spoofing Detector}
\label{sec:pd}

\begin{figure}[t]
    \centering
    \includegraphics[width=0.45\textwidth]{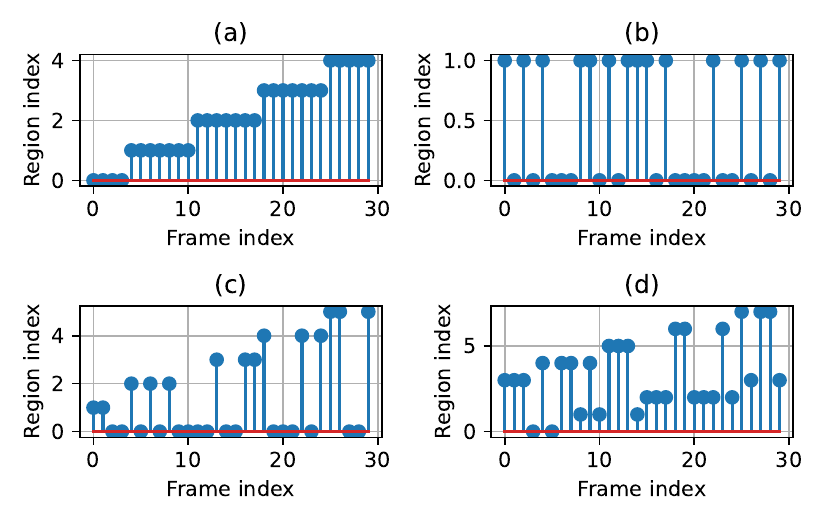}
    \caption{Examples of region sequences. (a) Single moving transmitter. (b) Two static transmitters. (c) Two transmitters, one is static and the other moves. (d) Two moving transmitters.}
    \label{fig:regions}
\end{figure}

\begin{bullets}
    \blt[overview]The PCD is applied to all pairs of frames to obtain the
    decisions $\Dec{\IndFra}{\IndFra'}\define
        \DecPCD(\EstVecRss\NotFra{\IndFra},\EstVecRss\NotFra{\IndFra'})$.
    The question is how to combine these     decisions to choose between   $\HzSD$
    and $\HoSD$.
    \blt[overview]To this end, this section  proposes a decision rule $\DecSD:\rfield^{\NumFea\times
            \NumFra}\rightarrow \{\HzSD,\HoSD\}$ based on the notion of
    \emph{region}. For didactical purposes, this concept is first
    introduced in an ideal scenario.

    \blt[region]
    A \emph{region} $\Reg\subset\{0,\ldots,\NumFra-1\}$ is a \emph{set of
        indices} such that $\Dec{\IndFra}{\IndFra'}=\HzPCD$ for all
    $\IndFra,\IndFra'\in \Reg$ whereas $\Dec{\IndFra}{\IndFra'}=\HoPCD$ if
    $\IndFra\in \Reg$ and $\IndFra'\notin \Reg$ (or vice versa since
    $\Dec{\IndFra}{\IndFra'}=\Dec{\IndFra'}{\IndFra}~ \forall \IndFra,\IndFra'
    $). Note that, as per this definition, this is not  a geometric notion. However, the term \emph{region} was chosen since one can intuitively identify  each region with a spatial area where the channel
    is roughly constant to the eyes of $\DecPCD$.

    \blt[Region sequences]%
    \begin{bullets}%
        \blt[assumption]With this in mind, consider the following
        assumption:
        \begin{align}
            \nonumber
            \text{(As2)~The frame sequence can be partitioned into regions.}
        \end{align}
        In other words, (As2) states that, for each frame sequence $\EstMatRss$, there is a
        collection of regions such that  each frame $\EstVecRss[\IndFra]$ belongs to one and only one
        region.
        \blt[approx true]However, the above definition of region is too strict
        for (As2) to hold in general, but analyzing  simple cases
        offers insight that will be used to relax the notion of region so that (As2)
        holds.


        \begin{bullets}%
            \blt[single static user]Consider first a single
            legitimate user that remains static and assume that all PCD decisions are correct. Since
            $\Dec{\IndFra}{\IndFra'}=\HzPCD$ for all
            $\IndFra,\IndFra'$,   all frames belong to the same
            region $\Reg=\{0,\ldots,\NumFra-1\}$.
            \blt[two static users]With two static users, one being the attacker, $\Dec{\IndFra}{\IndFra'}=\HzPCD$
            if $\IndFra,\IndFra'\in \SetIndLeg$ or $\IndFra,\IndFra'\in \SetIndAtt$, and
            $\Dec{\IndFra}{\IndFra'}=\HoPCD$ otherwise.
            Thus, one can set $\Reg_1=\SetIndLeg$ and $\Reg_2=\SetIndAtt$.

            \blt[single moving user]Consider now a single user  at
            one location up to frame $\IndFra_0$ and at another location from frame $\IndFra_0+1$ on. Then
            $\Reg_1=\{0,\ldots,\IndFra_0\}$ and
            $\Reg_2=\{\IndFra_0+1,\ldots,\NumFra-1\}$. This observation, which is extensible to an
            arbitrary number of spatial locations, constitutes a discretization
            of the user movement but can be made arbitrarily accurate by
            increasing the number of considered points and its impact will be
            mitigated once the definition of region is re`laxed.
            \blt[two moving users]The cases of two moving  users or a static
            user and a moving user can be handled similarly.

            \blt[examples]To  illustrate these cases,
            Fig.~\ref{fig:regions} shows examples of \emph{region sequences}, namely
            \blt[def seq]
            sequences of
            indices $\IndReg[0],\ldots,\IndReg[\NumFra-1]$ where $\IndReg[\IndFra]$
            is the index of the region to which frame $\IndFra$ belongs, i.e.,
            $\IndFra\in \Reg_{\IndReg[\IndFra]}~\forall \IndFra$. Note that
            region indices are arbitrary.

            \blt[conclusion]While (As2) holds in  the above ideal scenarios, it
            will not generally hold. The next section copes with this issue.


        \end{bullets}%

    \end{bullets}%

    \blt[rest of section overview]



\end{bullets}

\subsubsection{Region Estimation via Community Detection}
\label{sec:community}

\begin{bullets}%
    \blt[not transitive]The reason why it is not generally possible to decompose
    the frame sequence as a disjoint union of  regions is that $\DecPCD$ is not
    generally \emph{transitive}, which means that even though
    $\DecPCD(\EstVecRss,\EstVecRss')= \DecPCD(\EstVecRss',\EstVecRss'')=\HzPCD$,
    it may happen that $\DecPCD(\EstVecRss,\EstVecRss'')=\HoPCD$. This may be
    caused by noise or other factors. However, given (As1), one expects that
    transitivity is violated only for a few frames, a fact exploited next.
    %

    \blt[graph] Consider a graph $\Gra$ where each
    node corresponds to a frame and there is an edge between  node $\IndFra$ and
    node $\IndFra'$ iff $\Dec{\IndFra}{\IndFra'}=\HzPCD$. A decomposition of a
    frame sequence into regions induces a decomposition of the graph into
    \emph{fully-connected components}. Recall that the latter comprise sets of nodes that are
    connected to each other but not to any other node outside the
    component~\cite[Ch. 2]{kolaczyck2009}.
    %
    %
    \blt[community] This notion  can be relaxed into that of
    \emph{community}. A community  informally refers to a set of nodes
    with a large number of edges between them and few edges to nodes outside the
    community~\cite[Sec. 4.3.3]{kolaczyck2009}. The term is coined in analogy to a
    community in a social
    network,  e.g.,  a group of friends.
    \blt[community detection]Algorithms to determine the
    communities of a graph abound and are referred to as \emph{community
        detection algorithms}. A popular one is the \emph{Louvain
        method}, which optimizes a metric called
    \emph{modularity}~\cite{blondel2008unfolding}.

    \blt[proposed]
    The proposed method  takes $\EstMatRss$ as input and
    constructs the adjacency matrix of $\Gra$ by evaluating $\DecPCD$ for all
    $\EstVecRss[\IndFra]$ and $\EstVecRss[\IndFra']$ with $\IndFra<\IndFra'$.
    The Louvain method is then applied to partition the graph into
    communities, which yields a sequence
    $\IndReg[0],\ldots,\IndReg[\NumFra-1]$ with the indices of the community
    corresponding to each frame. This sequence will still be referred to as a
    region sequence although, strictly speaking, each community does not exactly
    correspond to a region according to the definition of region  at the beginning of Sec.~\ref{sec:pd}. However, it may be identified with a region in a softer sense,
    since a community may miss  a few edges inside or may have edges
    to nodes outside.

\end{bullets}%

\subsubsection{Decision Rule}
\label{sec:rule}
\begin{bullets}%
    \blt[overview] Having determined
    $\IndReg[0],\ldots,\IndReg[\NumFra-1]$, the final step is to decide $\HzSD$
    or $\HoSD$.
    \blt[test]This process comprises two steps: first, a test
    statistic $\Sta$ is evaluated. Then, this statistic is compared with a
    threshold $\ThrSD$ to decide $\HoSD$ if $\Sta(\IndReg[0],\ldots,\IndReg[\NumFra-1])>\ThrSD$ and $\HzSD$ otherwise.
    Thus, it remains only to find a suitable $\Sta$.

    \blt[statistic]
    \begin{bullets}%
        \blt[derivation]

        Clearly, $\Sta$
        should generally take larger values under $\HoSD$
        than under $\HzSD$. Since a sequence with no transitions ($\IndReg[\IndFra]=\IndReg[\IndFra']$ for all $\IndFra,\IndFra'$) would
        likely correspond to $\HzSD$, it follows that $\Sta$ must be small in this case. On the other hand, transitions (i.e., indices $\IndFra$
        such that $\IndReg[\IndFra]\neq \IndReg[\IndFra+1]$) indicate either
        movement or the presence of an attack. Thus, a natural  $\Sta$ is
        the number of transitions likely caused by an attack. To
        distinguish both kinds of transitions, observe from
        Fig.~\ref{fig:regions} that movement generally results in transitions to
        regions  not visited previously, whereas
        attack-related transitions revisit regions.

        \blt[definition]These considerations motivate setting $\Sta$ to be the number of transitions to a previously visited region. Specifically, one can write
        the test
        statistic as
        \begin{align}
            \label{eq:stacount}
            \Sta(\IndReg[0],\ldots,\IndReg[\NumFra-1])=
            \sum_{\IndFra=3}^{\NumFra-1}
            \Wei[\IndFra],
        \end{align}
        where
        $\Wei[\IndFra] = 1$ if $\IndReg[\IndFra]\neq \IndReg[\IndFra-1]$ and there exists a $ \IndFra'<\IndFra$ such that $\IndReg[\IndFra]=\IndReg[\IndFra']$, and $\Wei[\IndFra] = 0$ otherwise.

        \blt[weighted metric] However, due to noise and wrong decisions of
        $\DecPCD$, it is desirable to confer a greater weight to those terms in \eqref{eq:stacount} that correspond to
        transitions to a \emph{recently} visited region. In particular, if
        $|\cdot|$ denotes cardinality, then the number of distinct\footnote{Recall that, due to the  definition of \emph{set}, duplicates count as~one.} regions visited before returning to $\IndReg[\IndFra]$ is
        $\NumRegCom[\IndFra]\define
            |\{
            \IndReg[\IndFra^*+1],\IndReg[\IndFra^*+2],\ldots,\IndReg[\IndFra-1]
            \}|$, where
        $\IndFra^*$ is the largest index such that $\IndFra^*< \IndFra$ and  $\IndReg[\IndFra^*]=\IndReg[\IndFra]$.
        Since an attack is likely to result in low values of $\NumRegCom[\IndFra]$,
        it makes sense to replace  $\Wei[\IndFra]$ in \eqref{eq:stacount} with $\Wei[\IndFra]=1/\NumRegCom[\IndFra]$ if $\NumRegCom[\IndFra]\neq 0$ and $\Wei[\IndFra]=0$ otherwise.


        \blt[threshold] The threshold $\ThrSD$ is then  set e.g. to satisfy a
        target probability of false alarm  or to minimize the
        probability of error.

    \end{bullets}%
\end{bullets}%

\section{Experiments}
\label{sec:experiments}

\cmt{Overview} This section empirically validates the proposed
algorithm using real data. A link to the code and
data is provided on the first page.

\cmt{simulation setup}%
\begin{bullets}%
    \blt[data collection]%
    \begin{bullets}%
        \blt[environment]Several datasets were collected indoors on the campus
        of the University of California, San Diego.
        \blt[transmitter] This section focuses on one
        of those datasets, where a transmitter was sequentially placed at 52 locations along
        4 parallel lines. For the sake of reproducibility, the transmitted
        signal $\TxSig(t)$ is a 5 MHz sinusoid at a  2.3
        GHz carrier frequency. \blt[receivers]For every transmitter location, each of the four receivers with four
        antennas  record 4888 samples of the received signal. Thus,
        $\NumFea$ can be set between 1 and~4$\cdot$4=16 by selecting a subset of
        the receivers and antennas.

    \end{bullets}


    %
    \blt[tested algorithms]The PCD proposed in
    Sec.~\ref{sec:pcd}, referred to as \emph{DNN-based classifier} (DNNC), is benchmarked with three other PCD algorithms.
    \begin{bullets}%
        \blt[DBC]The first two, termed distance-based classifiers (DBCs),
        decide $\HoPCD$ iff $\| \EstVecRss-\EstVecRss'\|_q>\ThrPCD$, where $q$ is 2
        for DBC($\ell_2$) and 1 for DBC($\ell_1$). The threshold $\ThrPCD$ is
        adjusted to maximize the accuracy over the training pairs.
        \blt[KMC]Following
        \cite{chen2007detecting,yang2012detection}, the third benchmark is a
        K-means classifier that clusters the training vectors
        $\EstVecRss\NotEst{\IndEst}(\Loc_{\IndDat})$ into $\NumClu$
        clusters. Then it applies the same rule as DBC($\ell_2$) over the two
        feature vectors that result from collecting the $\NumClu$
        Euclidean distances from the given $\EstVecRss$ and $\EstVecRss'$ to
        all centroids.

    \end{bullets}

\end{bullets}

\subsection{Position Change Detection}
\cmt{overview} This section analyzes  performance when solving
\eqref{eq:test}. \cmt{performance metrics}To this end, the adopted metric is the
accuracy measured on a test data set comprising the $52-\NumDat$ locations that
are not used for training and validation. Recall from Sec.~\ref{sec:training} that, among these $\NumDat$ locations, $\NumDat\Tra$ are used for training and $\NumDat\Val$ for validation.
To estimate the accuracy of each algorithm, a new subset of $\NumDat$ locations
among all 52 available locations is chosen at each Monte Carlo iteration. Also, the DNN weights
are randomly initialized at each iteration.

\cmt{description of the experiments}
\begin{bullets}
    \blt[exp. 1]To investigate the
    number $\NumDat$ of training locations that are  needed to
    attain a reasonable accuracy,  Fig.~\ref{fig:trainingpositions} (top) depicts
    accuracy vs.\  $\NumDat$. It is observed that (i) the accuracy of DNNC is reasonably high
    with just 45 locations, (ii) DNNC learns much more than the benchmarks
    with new data. \cbr{The horizontal axis begins at $\NumDat=10$ because DNNC
        needs to split the given $\NumDat$ positions into  training and
        validation positions and, therefore, using $\NumDat<10$ results in
        noisy estimates of the training and validation losses;
        cf.~\cite[Ch. 2]{cherkassky2007}.}

    \begin{figure}[t]
        \centering
        \includegraphics[width=0.5\textwidth]{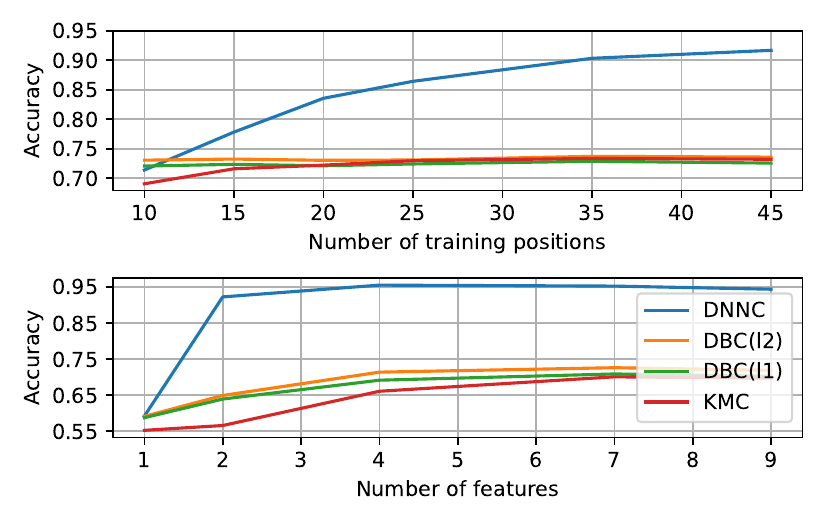}
        \caption{[Top] Accuracy vs. number of training positions $\NumDat$ for the
            proposed algorithm (DNNC) and the competing algorithms ($\NumSam=16$,
            $\NumFea=16$,
            $\NumPai\Tra=1250$,  $\NumPai\Val=150$, $\NumDat\Tra=0.8\NumDat$,
            $\NumClu=15$).   [Bottom] Accuracy vs. number of features $\NumFea$ for DNNC and the competing algorithms ($\NumSam=16$,
            $\NumPai\Tra=1250$, $\NumPai\Val=150$, $\NumDat=40$,
            $\NumDat\Tra=0.8\NumDat$, $\NumClu=15$).
        }
        \label{fig:trainingpositions}
        \label{fig:features}
    \end{figure}


    \blt[exp. 2]The second experiment assesses the impact  of the number
    of features. In this data set, the first 4
    features correspond to the first receiver, the second 4 features to
    the second receiver, and so on.  Fig.~\ref{fig:features} (bottom) shows that
    two features already result in  high accuracy for
    DNNC. Remarkably, this effect can be seen to be milder when the
    selected features are obtained by different receivers, suggesting
    that there is more information in the joint distribution of the RSS
    estimates from nearly-located antennas than from
    distant antennas. \cbr{Likely due to the fact that
        the latter are ``less synchronized'' than the former.}


    \blt[samples]In both experiments, only $\NumSam=16$
    samples are used, which corroborates that the goal of DNNC to provide  high
    temporal resolution was achieved. Further experiments omitted here show that the accuracy of the benchmarks decreases more
    abruptly than the one of DNNC if $\NumSam$ is reduced below 16.

\end{bullets}

\subsection{Spoofing Detection}

\begin{figure}[t]
    \centering
    \includegraphics[width=0.45\textwidth]{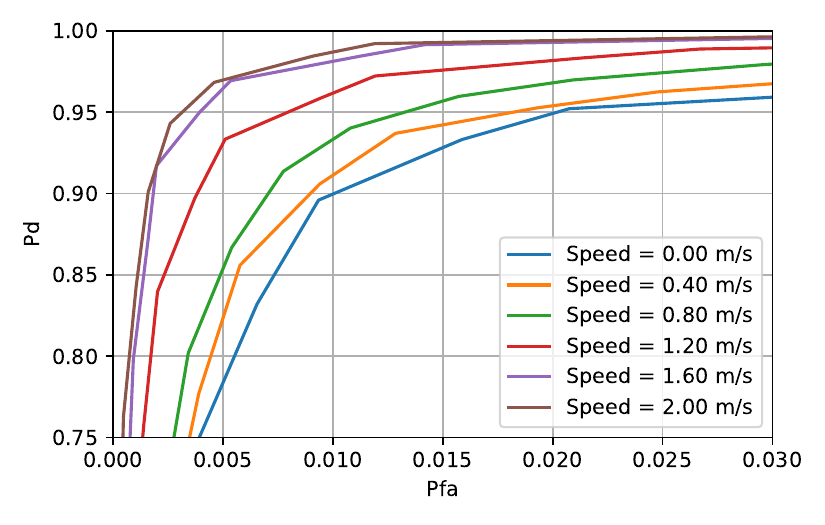}
    \caption{ROC curves for the proposed SD detector (100 frames/s, $\NumFra=30$, $\NumSam=16$,
        $\NumFea=16$,
        $\NumPai\Tra=3000$,  $\NumPai\Val=1000$, $\NumDat\Tra=0.8\NumDat$). }
    \label{fig:rocs}
\end{figure}

\begin{figure}[t]
    \centering
    \includegraphics[width=0.45\textwidth]{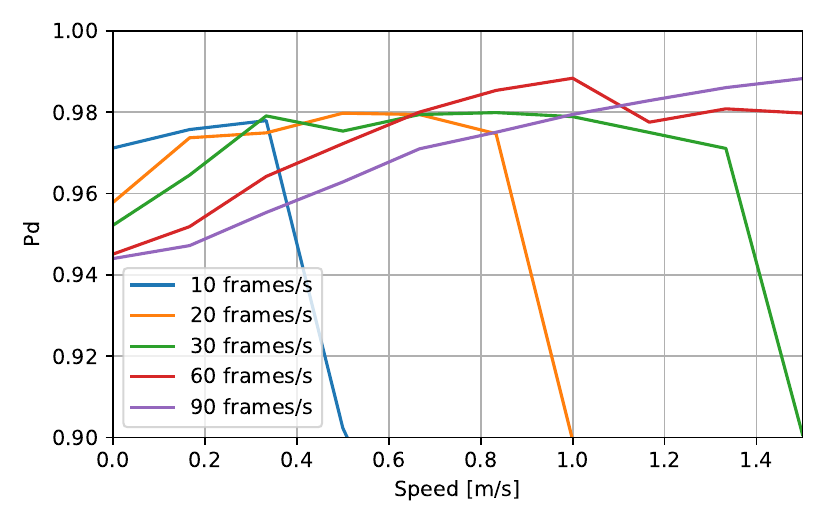}
    \caption{Probability of detection for a fixed probability of false alarm
        $\pfa$ as a function of the speed of the users ($\pfa=0.02$,
        $\NumFra=20$, $\NumSam=16$, $\NumFea=16$, $\NumPai\Tra=3000$,
        $\NumPai\Val=1000$, $\NumDat\Tra=0.8\NumDat$).}
    \label{fig:PdVsSpeed}
\end{figure}

\begin{figure}[t]
    \centering
    \includegraphics[width=0.45\textwidth]{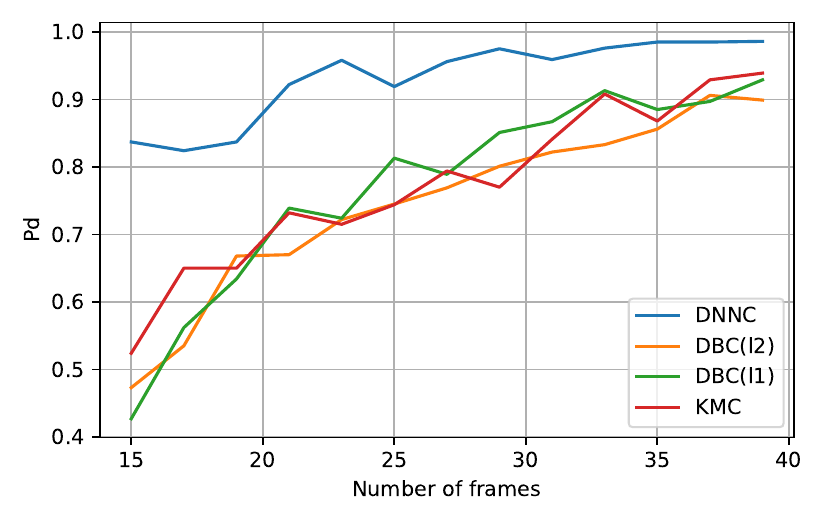}
    \caption{Probability of detection of spoofing attacks vs.  $\NumFra$ for different PCDs (10 frames/s, $\pfa=0.02$, $\NumSam=16$, $\NumFea=16$,
        $\NumPai\Tra=3000$, $\NumPai\Val=1000$, $\NumDat\Tra=0.8\NumDat$). }
    \label{fig:PCDcomparison}
\end{figure}

\cmt{overview} This section studies performance when solving \eqref{eq:maintest}.
\cmt{data generation}At each Monte Carlo iteration, a frame sequence is generated.
\begin{bullets}%
    \blt[H0]Under $\HzSD$, the trajectory $\Loc(t)$  of the (single) user is
    generated as follows. First, obtain the time duration of the frame sequence,
    given by $\NumFra\RatFra$, where $\RatFra$ is the number of frames per second. The length $\Delta\Loc$ of the
    trajectory is therefore $\Delta\Loc=\NumFra\RatFra\Spe$, where $\Spe$ is the  speed of the user. Then, select one of the lines indicated at the beginning of
    Sec.~\ref{sec:experiments} uniformly at random and choose a point $\LocLin$
    on that line that is at least $\Delta\Loc$ from the first and last
    measurement locations on that line. The trajectory is therefore
    $\Loc(t)=\LocLin + \VecDir \Spe t$, where $\VecDir$ is one of the two unit vectors on the selected line chosen uniformly at random. Each
    $\EstVecRss[\IndFra]$ is  obtained using samples in the data set that correspond to  the
    measurement location that lies closest to $\Loc(\IndFra/\RatFra)$.

    \blt[H1] Under $\HoSD$, the frame sequences of both users are also generated following this procedure. Then, the frame sequence
    $\EstVecRss_1[0],\ldots,\EstVecRss_1[\NumFra-1]$ of user-1 is merged with the
    frame sequence $\EstVecRss_2[0],\ldots,\EstVecRss_2[\NumFra-1]$ of user-2 into a
    sequence $\EstVecRss[0],\ldots,\EstVecRss[\NumFra-1]$ where either
    $\EstVecRss[\IndFra]=\EstVecRss_1[\IndFra]$  or
    $\EstVecRss[\IndFra]=\EstVecRss_2[\IndFra]$, both with probability 1/2 and independently
    along $\IndFra$.

\end{bullets}%

\cmt{experiments}
\begin{bullets}%
    \blt[ROC] Fig.~\ref{fig:rocs} presents the
    \emph{receiver operating characteristic} (ROC) curves~\cite[Ch. 3]{kay2} for
    the test in \eqref{eq:maintest} when the SD from Sec.~\ref{sec:pd} is used.
    Observe the large   probability of detection ($\pd$) that is
    attained with low probability of false alarm ($\pfa$). Interestingly, the
    detection performance seems to improve with the user speed. This is
    because speed reduces the size of the communities, which facilitates
    the task of the community detection step.
    However, intuition suggests that this trend cannot  continue indefinitely since, at some point, assumption
    (As1) will cease to hold.

    \blt[Pd vs speed]To investigate this effect further,
    Fig.~\ref{fig:PdVsSpeed} depicts $\pd$ for a given $\pfa$ vs. speed. For each frame rate $\RatFra$, there is a \emph{critical}
    speed above which the probability of detection abruptly falls. This
    effectively constitutes a sharp transition between the regime where (As1) holds
    and the regime where it does not. Below the critical speed, $\pd$ attains a maximum  for a speed that depends on the frame
    rate.

    \blt[PCD comparison]The experiment in Fig.~\ref{fig:PCDcomparison},
    investigates the influence of the adopted PCD on the spoofing detection performance.
    The SD algorithm from Sec.~\ref{sec:pd} is used with the PCD
    algorithm from Sec.~\ref{sec:pcd} and with the benchmarks. As expected, the
    better performance of DNNC translates into a better performance in the spoofing detection
    task.

\end{bullets}%

\section{Conclusions }
\label{sec:conclusions}
\begin{bullets}
    \blt[conclusions]This work considers the detection of spoofing attacks in
    wireless networks.  RSS features were used since they require no
    synchronization among receivers and no knowledge about the communication
    protocol or standard. Due to the high spatial variability of multipath, the
    proposed detector relies on lower level decisions provided by a position-change
    detection algorithm. To account for errors in the latter, an algorithm for
    community detection on graphs is invoked. The sequence of detected
    communities is then used to decide if  a spoofing attack is taking place. To attain
    high spatio-temporal resolution, a position-change detector was developed  based on a
    DNN. Real data demostrated that the proposed scheme attains high $\pd$ and
    low $\pfa$ even for small training data sets. Remarkably, acquiring such
    data is   simple since the measurement locations need not be estimated or recorded.

\end{bullets}

\balance
\small
\printmybibliography
\end{document}